\documentstyle[aps,epsfig,multicol,fancyheadings]{revtex}
\begin{document}

\title{Failure regimes in (1+) dimensions in fibrous materials}

\author{I.~L.\ Menezes-Sobrinho$^1$, A.~T.\ Bernardes$^2$, 
and J.~G.\ Moreira$^3$}

\address{$^1$Departamento de F\'{\i}sica, Universidade Federal de Vi\c cosa,
36570-000 Vi\c cosa, MG, Brazil} \date{\today}
\address{$^2$Departamento de F\'{\i}sica, Universidade Federal de Ouro Preto,
35400-000 Ouro Preto, MG, Brazil} \date{\today}
\address{$^3$Departamento de F\'{\i}sica, Instituto de Ci\^encias Exatas,
Universidade Federal de Minas Gerais, C.P 702, 30123-970 Belo Horizonte, MG,
Brazil} \date{\today}

\maketitle

\begin{abstract}
In this paper, we introduce a model for fracture in fibrous materials
that takes into account the rupture height of the fibers, in contrast with
previous models. Thus, we obtain the profile of the fracture
and calculate its roughness, defined as the variance around the
mean height. We investigate the relationship between the fracture roughness
and the fracture toughness. 
\end{abstract}

\noindent {\it PACS \# \ \ 62.20.Mk, 64.60.Fe, 05.40.-a}

\begin{multicols}{2}[]
\narrowtext

\section{Introduction}\label{intro}
The fracture of a material under different loading conditions is 
a  complicated phenomenon that continues to attract the attention
of several physicists and engineers. The fracture follows a sequential process
of  nucleation, growth, and coalescence of numerous cracks. These cracks tend
to create a rough profile which often can be described by self-affine scaling
\cite{mand}. In the last decade, great theoretical and experimental efforts
have been done trying to understand the process of formation and
pro\-pa\-gation of cracks in materials. A great number of studies have been
devoted to this process from the  point of view of statistical mechanics, that
utilize concepts such as percolation, fractals, scaling law,
etc. \cite{mand,livro}. A variety of computational models for material fracture
have been developed in recent years and provided interesting results
\cite{livro,sornete,al}.  However, the high degree of correlations between the 
constituents leads to a high computational cost. An alternative is to employ
fiber bundle models introduced more than forty years ago \cite{dan}, in which
bundles of unidirectional fibers form a system with low degree of correlations
allowing the fracture process to be simulated in a large scale. 

The fracture of the material can occur
through a big crack which percolates the sample. In this case we say that
the fracture is catastrophic. When the fusion of small
cracks cause the fracture of the material, it is called shredding \cite
{isma}. In the catastrophic regime the fracture profile is reasonably smooth,
while in the shredding regime it is very rough. 
A parameter of easy physical interpretation
used to characterize fracture surfaces is the roughness. It is defined 
as the variance around of the mean height of the fracture profile, 
and measures the complexity of the crack path. Thus, the rougher is the
fracture profile the harder is the crack path. The roughness has a 
direct relation to the fractal dimension, which characterizes the fractal 
character of the surface fracture. Therefore, a high fractal dimension 
indicates a very rough fracture profile \cite{mand,jose}. 
 
Another important parameter in the study of fracture, is the fracture
toughness.  This quantity measures the amount of energy that a material
can absorb before it fractures. The fracture toughness is intimately related
to the amount of cracks that appear in the material \cite{livro}.
Therefore, the larger the number of cracks in the sample more energy will be
absorbed before catastrophic fracture occurs. There are several works in the
literature which investigate the relationship between the fracture roughness
and the fracture toughness \cite{jose,shou,pez,mu,tp}. 

In the present Rapid Communication, we simulated a model for fracture in
fibrous materials which allows us to obtain the fracture profile of samples in
contrast with previous mo\-dels \cite{isma,isma1,zhang,rava,ze,zh}. We
calculate the profile roughness and used the obtained values to define the
transition between two fracture regimes: catastrophic and shredding. We also
investigate the relationship between the fracture roughness and toughness for
fibrous materials.

\section{Model}\label{num}
Our model consists of a (1+1)d bundle of $N_0$ parallel fibers
all with the same elastic constant $k$. In order to simulate the height
of the sample, the fibers are divided in $\eta$ segments with the same length. 
The fiber bundle is fixed at both ends to two parallel
plates. One  plate is fixed, and in the
other, a constant force $F$ is applied, for example, by hanging a weight on
it. This force is shared in the same amount, $\sigma$, on each fiber of the
bundle, which undergo the same linear deformation $z=F/Nk$, where
$N$ is the number of unbroken fibers. When the deformation $z$  reaches a
critical value $z_c$, the failure probability of an isolated fiber is equal to
1. The failure probability  of a fiber {\it i} is given by \cite{isma} 
\begin{equation}
P_i(\delta,t)={\delta\over(n_i+1)}\exp\left[{(\delta^2-1)\over t}\right],
\label{eq1} \end{equation}
where $n_i$ is the number of unbroken neighboring fibers,
$\delta=z/z_c=F/Nkz_c$ is  the strain of the material, $t=K_BT/E_c$ is the
normalized temperature, $K_B$ is the Boltzmann constant, $T$ is the
absolute temperature, and $E_c$ is the critical elastic energy. In this model,
besides finding the failure probability of a fiber, we have to indicate in
which segment it breaks. This segment is randomly selected and the
probability of the fiber to break in it is given by 
\begin{equation}
\phi_j(m_j)={(m_j+1)\over \zeta},
\label{eq2}
\end{equation}
where $m_j$ is a vector which indicates how many times a segment $j$
broke and $\zeta=\sum_j (m_j+1)$. The form of the Eq.~(\ref{eq2}) simulates 
a concentration of tension near to the region where the fiber bundle is  
more weak.

At the beginning of the simulation, the bundle is submitted to a initial strain
given by 
\begin{equation}
\delta_0={z_o\over z_c}={F\over {N_0kz_c}}.
\label{eq3}
\end{equation}
 At each time step
of the simulation we randomly choose a fiber of a set of $N_q =qN_o$ unbroken
fibers. The number $q$ represents a percentage of fibers and allows
us to work with any system size. Then, using Eq.~(\ref{eq1}) we evaluate
the fiber failure probability $P_i$ and compare it with a random
number $r$ in the interval [0,1). If  $r<P_i$ the fiber breaks. We then choose
a segment $j$ in the fiber and evaluate its probability $\phi$ to break, 
using Eq.~(\ref{eq2}). If the  probability $\phi$ is
higher than the random number $\rho$ the fiber breaks in the chosen segment.
If not, we analyze the neighboring segments $(j+1)$ and $(j-1)$ and again,
evaluate the probability $\phi$. If the condition $\rho<\phi$ does not hold to
neither of the  neighboring segments, we return to the initial segment and test
the condition  $\rho<\phi$ for a new value of $\rho$. This process continues
until the  condition $\rho<\phi$ is true. Once defined the segment where the
fiber breaks, we begin to test all neighboring unbroken fibers. The first
segment tested in the neighboring unbroken fibers is the segment where the
previous fiber broke. The failure probability $P_i$ of these neighboring
fibers increases due to the decreasing of $n_i$ and a cascade of breaking
fibers may begin. This procedure describes the propagation of a crack through
the fiber bundle, which occurs in the perpendicular direction to the applied force. 
The process of propagation stops when the
test of the probability does not allow rupture of any other fiber on the
border of the crack or when the crack meets another already formed crack. The
same cascade propagation is attempted by choosing another fiber of the set
$N_q$. After all the $N_q$ fibers have been tested, the strain $\delta$ is
increased if some fibers have been broken. Since the force is fixed, the
greater the number of broken fibers, the larger is the strain on the intact
fibers and the higher is their failure probability. Then, another set of $N_q$
unbroken fibers is chosen and the entire rupture process is restarted. The
simulation terminates when all the fibers of the bundle are broken, i.e.,
when the bundle breaks apart. At this stage we consider the profile of the
fracture and analyze its statistical properties.

\section{Results}\label{scale}

Using the model described in the preceding section we simulated the fracture
of a fibrous materials under a static force $F$. In these simulations we
considered the elastic constant $k=1$, the critical deformation $z_c=1$, the
number of segments $\eta=100$ and the normalized temperature $t=0.5$.

\begin{figure}[f]
\centerline{\epsfig{file=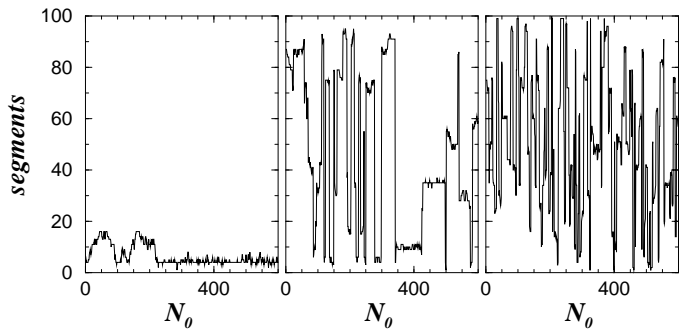,width=9cm,angle=-0}}
\caption{Fracture profiles for three different forces. From left to right
we have: $F=4000$, $F=2800$ and $F=2000$ arbitrary unity. In this
simulation we used a total of 600 fibers.}  
\label{fig1}  
\end{figure}

Figure \ref{fig1}
shows the fracture profile for a set of $N_0=600$ fibers and three different
forces. Notice that, the lower the applied force the larger is the roughness of
the fracture profile, i.e., more irregular is the crack path. For
$F=4000$ only one crack propagates in the material and for $F=2800$ and
$F=2000$ more than one crack crosses the material. It can be seen in
Fig. \ref{fig1}, that the rupture of the sample begins in different segments.
This  occurs because we did not consider a deterministic starting notch in
our simulations. 

\begin{figure}[f]
\centerline{\epsfig{file=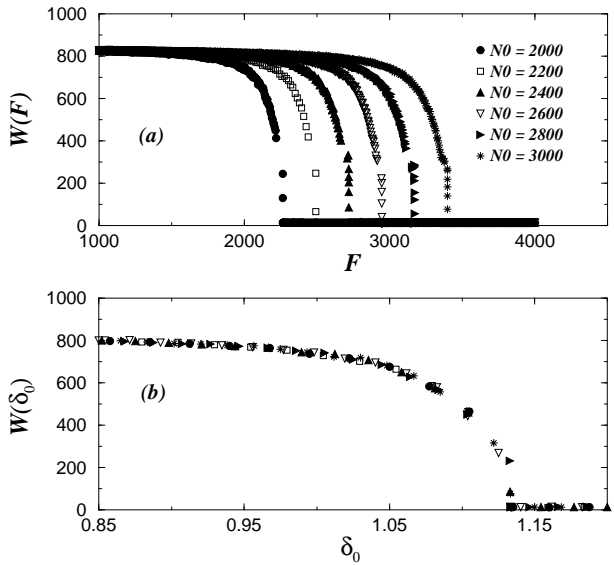,width=8cm,angle=-0}}
\caption{ (a) Roughness $W(F)$ as function of the applied force
$F$ for different number of fibers $N_0$. The data were averaged over
1000 samples. (b) The x-axis was converted to the initial strain
$\delta_0 = F /N_0$. } 
\label{fig2} 
\end{figure}

Figure \ref{fig2}(a) shows the results obtained for the
roughness $W$ as a function of the applied force $F$ for different
values of $N_0$. To measure the roughness of the fracture profile we used
the method of the best linear least-square fitting described in Ref.
\cite{jaf}. In this plot we can observe that the roughness $W$ decreases as
the force $F$ increases, and after a critical force $F_c$ it stays constant.
For each system size we have a characteristic value for $F_c$, which
increases with the increase of the number of fibers $N_0$. This indicates that
the greater the number of fibers in the bundle, the tougher is the sample. For
$F >F_c$ the fracture is catastrophic, i.e., the breakage of a
fiber induces the rupture of the whole bundle. In this case the bundle breaks
with only one crack. In this region the crack propagates in the material
with high speed, leading to a quite rapid rupture process. For $F
<F_c$ the rupture of the bundle occurs due to the formation of small cracks,
which weaken the bundle. Here, the crack speed tends to zero and a slow
process of successive ruptures appears in the material.

In Fig. \ref{fig2}(b), the x axis was converted to the
initial strain $\delta_0=F/N_0$. The data are now found to collapse all 
on the same roughness-strain curve. From this curve we can find the critical
initial strain $\delta_{0c}$, which does not depend on the system size. Our
results indicate $\delta_{0c}=1.134$. We may assume that, for an initial
strain greater than $\delta_{0c}$ just one crack provokes the rupture of the
material and that, for a initial strain below
$\delta_{0c}$ the rupture occurs through the fusion of small cracks. 

The critical value $\delta_{0c}$ can be obtained analytically. The 
failure probability, Eq.~(\ref{eq1}),  can be written as 

\begin{equation}
P_i(\delta,t)={\Gamma (t,\delta)\over (n_i+1)},
\label{eq4}
\end{equation}
where $\Gamma (t,\delta)$ is defined as \cite {isma}

\begin{equation}
\Gamma(t,\delta)=\delta exp\left[{{\delta^2 -1}\over t}\right].
\label{eq5}
\end{equation}

Using Eq.~(\ref{eq4}) we
may  observe that for $\Gamma(t,\delta)=2.0$ the breakage of any fiber induces
the rupture of the bundle with just one crack. Then, we can assume that
there is a critical value for $\Gamma(t,\delta)=2.0$ that defines the
transition between two regimes. In the first regime, a catastrophic
fracture occurs in the first attempt to break the bundle, while in the second
one the rupture of the bundle occurs due to the formation of small cracks,
which weaken the bundle. From Eq.~(\ref{eq5}) we can show that the
normalized temperature can be given by  \begin{equation}
t={{\delta_{0c}^2-1}\over {ln(\Gamma_c)-ln(\delta_{0c})}}. \label{eq6}
\end{equation} So, for $t=0.5$ and $\Gamma_c(t,\delta)=2.0$, Eq. (\ref{eq5})
will be valid only if $\delta_{0c}\approx 1.134.$ 

Now we proceed to the evaluation of the fracture toughness $k_c$. It can be
defined by the work done to break the fiber bundle and is given by 

\begin{equation}
K_c=\sum_i \tau_i,
\end{equation} 
where $\tau_i$ is the work done to break each fiber of the bundle. The 
work $\tau_i$ is obtained by the following expression 

\begin{equation}
\tau_i={1\over 2}kz_i^2,
\end{equation}
where $z_i$ is the deformation of the fiber
{\it i}.

\begin{figure}[f]
\centerline{\epsfig{file=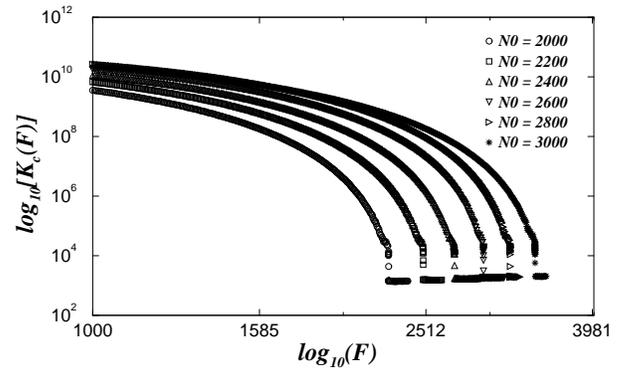,width=8cm,angle=-0}}
\caption{Fracture toughness $K_c$ vs the applied force $F$ for
different number of fibers $N_0$. The data were averaged over 1000
statistically independent samples.}  \label{fig3} 
\end{figure}

Figure \ref{fig3} shows the log-log plot of the fracture toughness $K_c$
versus the force $F$ for different system sizes. This figure shows
how the fracture toughness $k_c$ decreases with the increase of the force
$F$, until it reaches a minimum value. This value is attained when a
critical force $F_c$ is applied to the system. Above of $F_c$ the fiber
bundle breaks catastrophically. It is known experimentally that a
catastrophic break consumes little energy. As the force $F$ decreases
below $F_c$, the material can absorb more energy before it fractures and
the number of cracks in the material increases. Also, notice that the fracture
toughness increases with the system size. Thus, the higher the number of
fibers the more energy will be absorbed by fracture.

Some works in the literature conjecture that there is a relationship
between the fractal dimension and the fracture toughness \cite
{jose,shou,pez,mu}, while others conjecture that such relationship does not
exist \cite {bou,sch}. In this paper we investigated the connection between
the roughness $W$, directly related to the fractal
dimension, and the toughness fracture $K_c$. 

\begin{figure}[f]
\centerline{\epsfig{file=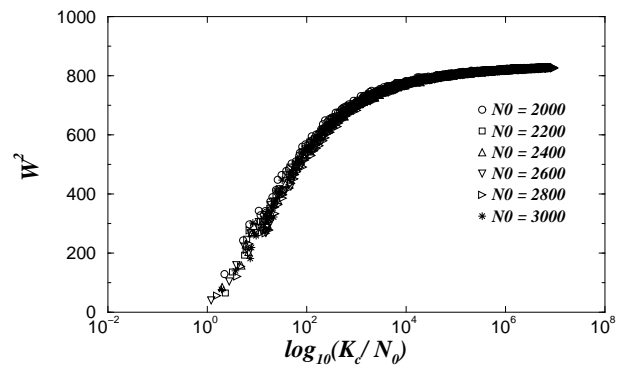,width=8cm,angle=-0}}
\caption{Roughness $W$ vs the fracture toughness $K_c$ for
different number of fibers $N_0$. The data were averaged over 1000
statistically independent samples.} 
\label{fig4}
\end{figure}

Figure \ref{fig4} shows how the roughness $W$ changes with the toughness
$K_c$ for several number of fibers $N_0$. In this fi\-gu\-re we can see that
the roughness $W$ increases with the increase of the fracture toughness
$K_c$, until it reaches a stationary state. We can interpret this
figure in the following manner: with the increase of $K_c$, small cracks
appear at all parts of the sample, making the fracture profile rougher. In the
saturation region, we have a larger number of small cracks and they are
totally uncorrelated. The phenomenon of saturation constitutes a finite size
effect and is related to the number of segments in which the fibers are
divided.

\section{Conclusion}\label{discussion}

In conclusion, we studied a model for fracture in fibrous materials in
(1+1)dimensions which take into account the rupture height of the fibers, in
contrast with previous models. We obtained the fracture profile and evaluate
its roughness and toughness. In this work we show that in the catastrophic
regime the roughness of the fracture profile is reasonably smooth. In the
shredding regime, in which slow cracks are formed in the material, the
fracture profile is very rough. In this regime the energy necessary to break
the material is higher than in the catastrophic regime.  Our results indicate
that the roughness $W$ is related to the fracture toughness. We believe that
the search for possible relationships between the roughness and the fracture
toughness could stimulate the further studies in order to check
whether these relations are valid or not. 

\medskip
\centerline{\bf Acknowledgments}
\medskip

\noindent
We thank Marcelo Lobato Martins and Marcos da Silva Couto for hel\-pful
cri\-ti\-cism of the manuscript. I.L.M. and A.T.B. acknowledge the kind
hospitality of the Departamento de F\'{\i}sica, UFMG. We also acknowledge the
FAPEMIG (Brazilian agencies) for financial support. 



\end{multicols}
\widetext
\end{document}